\documentstyle[prb,aps,epsfig,prbbib,amsmath,amssymb]{revtex}
\begin{document}

\draft
\twocolumn[\hsize\textwidth\columnwidth\hsize\csname  
@twocolumnfalse\endcsname
\title{Internal structure of preformed Cooper pairs}
\author{N. Andrenacci, H. Beck}
\address{ Institut de Physique, Universit\'e de Neuchatel, 
\\CH-2000 Neuchatel,  Switzerland}
\date{\today}
\maketitle
\hspace*{-0.25ex}

\begin{abstract}

In order to obtain information about the internal structure of the preformed pairs in the pseudogap state of high $T_c$ superconductors, we calculate
the propagator of 
 a singlet pair with center of mass coordinate ${{\mathbf{r}}}$, and relative
 distance
$\pmb{\rho}$, by solving the Bethe-Salpeter equation,
 representing the sum over repeated two-particle scattering events due to a distance dependent attraction.
We define then a  ``pair structure function'' $g_{P}({\mathbf{P}},\pmb{\rho}) $ that depends on the internal distance $\pmb{\rho}$ between the partners and on the momentum  ${\mathbf{P}}$ of the pair.  We calculate this function both for a local potential and $s-$wave symmetry of the order parameter and for a separable potential and  $d-$wave symmetry of the order parameter. The influence of the center of mass momentum, strenght of the interaction, temperature, density of particles and of the pseudogap in the one-electron spectrum is studied for both cases.
 
\end{abstract}
]
\narrowtext

\section{Introduction}

High-temperature cuprate superconductors show remarkable deviations from Fermi liquid behaviour in their 
normal state, in particular in the underdoped region. One of the most striking features is the opening of 
a pseudogap above the superconducting critical temperature $T_c$ and below another temperature, $T^*$, that increases when 
the doping is reduced\cite{Timusk}. This pseudogap appears to have the same angular dependence and magnitude as the 
superconducting gap below $T_c$, with which it seems to merge at the phase transition. In the pseudogap region, thermodynamic quantities and 
transport coefficients also deviate from Fermi liquid behaviour\cite{Randeira}.
To explain these anomalies, different scenarios can be invoked. One of them is based on the formation in the pseudogap region of incoherent Cooper pairs\cite{ZU}, also called ``preformed'' pairs\cite{emery}. At a lower temperature, phase coherence is established among these pairs leading 
to the superconducting transition. In this respect, the superconducting transition can be regarded almost as a Bose condensation of preformed Cooper pairs\cite{NSR}, while the BCS behaviour, in which the formation of the Cooper 
pairs and their condensation occur at the same temperature, is recovered as one approaches the 
overdoped region. 
The fact that the low temperature coherence length is much shorter in the underdoped regime is consistent 
with the picture of preformed pairs well localized in space.\cite{Uemura}
In this framework, it is interesting to investigate the structure of these preformed pairs. 

In the superconducting state, the Cooper pairs can be characterized by the so-called ``pair-size'' $\xi_{pair}$\cite{PistolesiStrinati}, which is obtained from the pair correlation function for particles paired with opposite spins. 
In the zero temperature limit, this reduces to the modulus square of the anomalous Green function. This definition can be generalized to be valid  in the pseudogap region, where the anomalous Green function is zero.

We base our calculations on the two-dimensionale attractive Hubbard model with an attraction acting either on electrons on the same lattice site or on neighbouring sites. Analytic and numerical work has shown that such a model can decribe the formation of electronic pairs of $s-$ wave and $d-$ wave, respectively, and the transition to the superconducting state. Going beyond the weak coupling limit, the formation af a pseudogap above the transition temperature is found. 
In order to describe the internal structure of preformed pairs in underdoped high-temperature superconductors we study the Green function for the operator which creates a pair 
of particles whose partners are at a certain relative distance, and study how this internal distance varies as a function of temperature, interaction strength and density of particles. To this end, we will consider 
a Bethe-Salpeter approach and introduce a $T-$matrix formalism, which takes into account the relevant interaction channel of repeated two-particle scattering.

In the next section we give an overview of our theoretical approach. In the case of a dilute system of charge carriers, which is of interest to us, the simplest approximation to the Bethe-Salpeter equation allows to separate the center of mass motion of a pair from its internal degree of freedom. The former is described by the standard two-particle $T-$matrix. The imaginary part of the latter is essentially the spectral function of a pair with momentum $\mathbf{P}$, but it does not give any information about the internal structure of the pair. The dependence of the pair propagator on the internal distance between the partners of the pair shows up in what we call the "pair structure function" $g_P(\mathbf{P},\pmb{\rho})$, which gives essentially the probability for the two partners of a pair, with a given center of mass momentum $\mathbf{P}$, of being at the distance $ \pmb{\rho}$ from each other. We present and discuss our numerical results in section \ref{sec:3}, where we show the influence of various parameters, such as temperature, chemical potential and interaction strength, on $g_P$ .The main conclusions from our work are summarized in section \ref{sec:4}. In Appendix \ref{app:A} we discuss the form of the one-electron spectral functions in the pseudogap region and in Appendix \ref{app:B} we relate our pair structure function to the total number of preformed pairs which is usually introduced in the framework of the Landau-Ginzburg description of the superconducting phase transition.

\section{Theoretical description of the pairs}

We consider an extended attractive two-dimensional Hubbard model given by the Hamiltonian:
\begin{eqnarray} \label{hamilt}
H&=&-\,t\,\sum_{<i,j>,\sigma}(c^{\dagger}_{i \sigma}c_{j \sigma}+\,h.c.)\,-\,\sum_{i,j}\,U(i-j)\, n_{i\downarrow}n_{j\uparrow} \nonumber \\
&-&\mu\,\sum_i\,(n_{i\downarrow}+n_{i\uparrow})
\end{eqnarray} 
where $t$ is the hopping parameter between nearest neighbor sites and $U(i-j)\,>\,0$ is an attractive interaction coupling the electronic density $n_{i\sigma}\,=\,c^{\dagger}_{i \sigma}c_{i \sigma}$ at two different sites $i$ and $j$. All our calculations are performed in the grand-canonical ensemble, using  the chemical potential $\mu$, rather than the electronic density, as a parameter. 

The Hubbard model has been analysed by many authors,
 using various kinds of approximation techniques, in order to describe the instability leading to superconductivity and to investigate the influence of the 
attractive interaction on the electronic properties, both above and below the 
superconducting transition temperature $T_c$. 
By increasing the strength of the attraction, it is possible to follow the evolution of a variety of physical quantities in a cross-over between a BCS-like 
behaviour to a scenario where superconductivity can be related to the condensation of preformed pairs.\cite{micnas} Our present objective consists in analysing, in the framework of (\ref{hamilt}), the structure of the 
preformed pairs in the normal 
state for $T\,>\,T_c$, in the domain where simulations show the formation of a pseudogap. 
One of the current interpretations of this unusual feature in the cuprates precisely invokes the presence of preformed electronic pairs, existing up to a temperature $T^*$, that can be 
substantially higher than $T_c$. The question of the structure of the pairs below $T_c$ will
 be treated elsewhere.

Let us consider the operator
\begin{equation}\label{2ptoper}
Q^{\dagger}({{\mathbf{r}}}_{ij},\pmb{\rho}_{ij})\,=\,c_{i \uparrow}^{\dagger}\,c_{j \downarrow}^{\dagger}\,,
\end{equation}
defined on the lattice sites $i$ and $j$ with coordinates $\mathbf{x}_i$ and $\mathbf{x}_j$, which creates a pair of electrons with opposite spins. We  define the center of mass coordinate of the pair  ${{\mathbf{r}}}_{ij}$ and the relative distance $\pmb{\rho}_{ij}$, as:
\begin{eqnarray}
{{{\mathbf{r}}}}_{ij}&=&\frac{1}{2}\,({\mathbf{x}}_i+{\mathbf{x}}_j) \nonumber \\
\pmb{\rho}_{ij}&=&{\mathbf{x}}_i-{\mathbf{x}}_j \nonumber \,.
\end{eqnarray}
Although $Q^{\dagger}$ and its adjoint do not satisfy the canonical commutation relations for
 independent particles, they can be used in order to describe electronic pair 
correlations through the dynamic correlation function:
\begin{eqnarray} \label{chifunc}
&&\chi({{{\mathbf{r}}}}_{ij},\pmb{\rho}_{ij},t;{{\mathbf{r}}}_{rs},\pmb{\rho}_{rs},t')\,=\,(-i)\, \\
&&\quad\times \,\langle\, T \left[ Q({{\mathbf{r}}}_{ij},\pmb{\rho}_{ij},t)Q^{\dagger}({{\mathbf{r}}}_{rs},\pmb{\rho}_{rs},t')\right] \,\rangle \,,\nonumber 
\end{eqnarray}
\noindent
where $T$ denotes the time ordering operator. 
Manifestly, $\chi$ is a two-electron Green's function $G_2$
\begin{equation} \label{2ptgreen}
G_2(1,2;1',2')\,=\,-\,\langle T\left[ c(1)c(2)c^{\dagger}(2')c^{\dagger}(1')\right]\rangle\,,
\end{equation}
with a special choice of arguments
(the numbers $i\,=\,1,\,2,\,1',\,2'$ are a shorthand notation for the generalized coordinates $\mathbf{x}_i, \, t_i, \, \sigma_i$ of space, time and spin  variables):
\begin{equation}\label{2ptfunc2ptoper}
\chi({{\mathbf{r}}}_{ij},\pmb{\rho}_{ij},t;{{\mathbf{r}}}_{rs},\pmb{\rho}_{rs},t')\,=\,i\,G_2(\mathbf{x}_i,t;\mathbf{x}_j,t;\mathbf{x}_r,t';\mathbf{x}_s,t')\,.
\end{equation}

A simple and convenient approximation for evaluating $G_2$ for a system described by the Hamiltonian (\ref{hamilt}) consists in solving the Bethe-Salpeter 
equation\cite{KB}:
\begin{eqnarray} \label{BS}
G_2(1,2;1',2')&=&G(1,1')G(2,2')\,-\,G(1,2')G(2,1')\nonumber \\
&+& i\, G(1,\underline{1})G(2,\underline{2})U(\underline{1}-\underline{2})G_2(1,2;\underline{1},\underline{2})
\end{eqnarray}
taking into account only repeated scattering events between two electrons (thus restricting our analysis to low density systems) and expressing $G_2$ in term of the ``full'' one-electron Green's function $G$ and of the attraction potential $U$. Here and in the following we use  imaginary time arguments, $0\,\leq\,t\,\leq i\beta,\,\beta\,=\,1/(k_BT)$, with the corresponding time-ordering. We also 
adopted in (\ref{BS}) the convention for which the bar under two identical arguments denotes 
integration over space and time. 

Introducing the special choice of variables needed for the pair correlation (\ref{2ptfunc2ptoper}) into the general Bethe-Salpeter equation (\ref{BS}) and denoting the dynamic correlation function (\ref{chifunc}) by $\chi({{\mathbf{r}}}_{ij}-{{\mathbf{r}}}_{rs},\pmb{\rho}_{ij},\pmb{\rho}_{rs}, t-t')\equiv \chi({{\mathbf{r}}}_{i},\pmb{\rho}_j,\pmb{\rho}_l, t)$ (where now ${{\mathbf{r}}}_{i}\,=\,{\mathbf{i}}a$ and $\pmb{\rho}_j\,=\,{\mathbf{j}}a$ are discrete vectors, that represent the distance among two sites  on a (square) lattice of constant $a$), one obtains
\begin{eqnarray}\label{BSChi}
\chi({{\mathbf{r}}}_{i},\pmb{\rho}_j,\pmb{\rho}_l,t)&=&-i \, {\mathcal{G}}({{\mathbf{r}}}_{i},\pmb{\rho}_j-\pmb{\rho}_l,t)\,+\,i\,\int\,dt' \sum_m \sum_k \nonumber \\
&\times & {\mathcal{G}}({{\mathbf{r}}}_{i}-{{\mathbf{r}}}_{m},\pmb{\rho}_j-{{\mathbf{r}}}_k,t-t')\,U({{\mathbf{r}}}_k)\, \nonumber \\
&\times& \chi({{\mathbf{r}}}_{m},{{\mathbf{r}}}_k,\pmb{\rho}_l,t')
\end{eqnarray}
where ${\mathcal{G}}$ is the particle-particle bubble:
\begin{equation}\label{gstorta}
{\mathcal{G}}({{\mathbf{r}}}_{i},\pmb{\rho}_j,t)\,=\,G({{\mathbf{r}}}_{i}+\frac{\pmb{\rho}_j}{2},t)\,G({{\mathbf{r}}}_{i}-\frac{\pmb{\rho}_j}{2},t)\,,
\end{equation}
\noindent
describing the motion of two independent particles whose properties have
 been renormalized by the interaction. The spin indices have been omitted, 
since we are interested in the particular configuration of opposite spin of the partners of the pairs, and the interaction is independent of the spin variable, so that no spin-flip processes have to be considered.

We now apply a Fourier transformation with respect to the variable ${{\mathbf{r}}}_{i}$
 - corresponding to the difference between the center of mass coordinates in 
$\chi$
- to equation (\ref{BSChi}), introducing the center of mass momentum ${\mathbf{P}}$, and we go from imaginary time to (bosonic) Matsubara frequencies $z_{\nu}$. This yields
\begin{eqnarray}\label{BSChimom}
\chi({\mathbf{P}},\pmb{\rho}_j,\pmb{\rho}_k,z_{\nu})&=&-\,i\,{\mathcal{G}}({\mathbf{P}},\pmb{\rho}_j-\pmb{\rho}_k,z_{\nu})\,+\,i\,\sum_i \, U({{\mathbf{r}}}_i)\nonumber \\
&\times& {\mathcal{G}}({\mathbf{P}},\pmb{\rho}_j-{{\mathbf{r}}}_i,z_{\nu})\,\chi({\mathbf{P}},{{\mathbf{r}}}_i,\pmb{\rho}_k,z_{\nu})\,.
\end{eqnarray}

For a local attraction $U({{\mathbf{r}}})\,=\,U_0\delta({{\mathbf{r}}})$ equation (\ref{BSChimom}) becomes algebraic and $\chi$ can immediately be related to the 
particle-particle bubble function ${\mathcal{G}}$ (see below). When the interaction has a finite spatial 
extension, which is needed for treating $d$-wave pairing, one has either to solve the 
integral equation (\ref{BSChimom}) or to introduce a separable form for the Fourier transform of $U$:
\begin{eqnarray} \label{utf}
\tilde{U}({\mathbf{k}}-{\mathbf{k}}')&\equiv&\sum_l \,e^{i({\mathbf{k}}-{\mathbf{k}}')\cdot\mathbf{x}_l}U(\mathbf{x}_l)\nonumber \\
&=&\sum_{\lambda}\,U_{\lambda}\phi_{\lambda}({\mathbf{k}})\phi^*_{\lambda}({\mathbf{k}})\,,
\end{eqnarray}
$\phi_{\lambda}$ being the various characteristic shape functions of the potential (the limit of the local potential can be retrieved by taking only the mode 
$\phi_{\lambda}\,=\,1$). In this more general case, equation (\ref{BSChimom}) 
is replaced by

\begin{eqnarray}\label{chisep}
\chi({\mathbf{P}},\pmb{\rho},\pmb{\rho}',z_{\nu})&=&-i\sum_{{\mathbf{k}}}\,e^{-i{\mathbf{k}}(\rho-\rho')}\tilde{{\mathcal{G}}}({\mathbf{P}},{\mathbf{k}};z_{\nu})+
i\sum_{\lambda}U_{\lambda} \nonumber \\
&\times& \sum_{{\mathbf{k}}}\,e^{-i{\mathbf{k}}\cdot\pmb{\rho}}\phi^*_{\lambda}({\mathbf{k}})\tilde{{\mathcal{G}}}({\mathbf{P}},{\mathbf{k}};z_{\nu})\nonumber \\
&\times & \sum_{{\mathbf{k}}'}\,e^{-i{\mathbf{k}}'\cdot\pmb{\rho}'}F_{\lambda}({\mathbf{P}},{\mathbf{k}};z_{\nu})\,,
\end{eqnarray}
where $\tilde{{\mathcal{G}}}({\mathbf{P}},{\mathbf{k}};z_{\nu})$ is the Fourier transform of the particle-particle bubble (\ref{gstorta}) with respect to the the center of mass position ${{\mathbf{r}}}_{ }$, internal distance $\pmb{\rho}$ and time, and $F_{\lambda}$ is an intermediate-step quantity obtained from the convolution 
of the Fourier transform  of the dynamic 
correlation function $\chi$ with respect to the variables $\pmb{\rho}_i$ and $\pmb{\rho}_j$, 
and the characteristic shape function of the potential:

\begin{eqnarray}\label{chitransf}
F_{\lambda}({\mathbf{P}},{\mathbf{k}};z_{\nu})&=&\sum_{{\mathbf{k}}'} \phi_{\lambda}({\mathbf{k}}')\,\sum_i \,\sum_j\,e^{i{\mathbf{k}}\cdot\pmb{\rho}_i}e^{-i{\mathbf{k}}'\cdot\pmb{\rho}_j}\nonumber \\
&\times & \chi({\mathbf{P}},\pmb{\rho}_i,\pmb{\rho}_j,z_{\nu})\,.
\end{eqnarray}

In the following, we will select one value of $\lambda$ and treat two cases separately:
\begin{itemize}
\item s-wave pairing: $\phi_{\lambda}({\mathbf{k}})\,=\,1$;
\item d-wave pairing: $\phi_{\lambda}({\mathbf{k}})\,=\,\cos{k_x}\,-\,\cos{k_y}$,
\end{itemize}
where  we have assumed the lattice constant to be $a=1$.

We will finally aim at the following quantities:
\begin{enumerate}
\item the usual two-particles T-matrix $T({\mathbf{P}},z_{\nu})$, which contains all the details of the interaction but does not show any dependence on the internal distance $\pmb{\rho}_i$ between the partners of the ``pair''. This quantity is given by

\begin{equation}\label{tmatrix}
T({\mathbf{P}},z_{\nu})\,=\,\frac{U_{\lambda}}{1\,-\,i\,U_{\lambda}\sum_{{\mathbf{k}}}|\phi_{\lambda}({\mathbf{k}})|^2\tilde{{\mathcal{G}}}({\mathbf{P}},{\mathbf{k}};z_{\nu})}\,;
\end{equation}

\item the spectral function $\Phi({\mathbf{P}},\pmb{\rho},\Omega)$ of a pair with center of mass momentum ${\mathbf{P}}$, energy $\Omega$ and internal distance $\pmb{\rho}$:

\begin{equation}\label{defPhi}
\Phi({\mathbf{P}},\pmb{\rho},\Omega)\,=\,Im{\{\chi({\mathbf{P}},\pmb{\rho},\pmb{\rho},\Omega+i0^+ )\}}\,;
\end{equation}

\item the ``pair structure function'' $g_P({\mathbf{P}},\pmb{\rho})$, that gives the internal structure of a pair having momentum ${\mathbf{P}}$ and internal distance $\pmb{\rho}$:

\begin{equation}\label{defdistr}
g_P({\mathbf{P}},\pmb{\rho})\,=\,\int \, d\Omega\, N_B(\Omega)\, \Phi({\mathbf{P}},\pmb{\rho},\Omega)\,,
\end{equation}
$N_B$ being the Bose-Einstein distribution function.

\end{enumerate}

The equations (\ref{chisep}) and (\ref{chitransf}) can now be solved in two steps. 
First, taking the double Fourier transform of (\ref{chisep}) one can relate $F$ to the Fourier transform of the ``bubble'' ${\mathcal{G}}$ with respect to the variable $\pmb{\rho}$:

\begin{equation}\label{fsol}
F_{\lambda}({\mathbf{P}},{\mathbf{k}};z_{\nu})\,=\,\phi_{\lambda}({\mathbf{k}})\frac{T({\mathbf{P}},z_{\nu})}{U_{\lambda}}\tilde{{\mathcal{G}}}({\mathbf{P}},{\mathbf{k}};z_{\nu})\,.
\end{equation}
Inserting this result into (\ref{chisep}) yields:

\begin{eqnarray}\label{chiTm}
\chi({\mathbf{P}},\pmb{\rho},\pmb{\rho},z_{\nu})&=&-i\,{\mathcal{G}}({\mathbf{P}},{\mathbf{0}},z_{\nu})\,+\,T({\mathbf{P}},z_{\nu})\nonumber \\
&\times& \hat{{\mathcal{G}}}({\mathbf{P}},\pmb{\rho},z_{\nu})\hat{{\mathcal{G}}}({\mathbf{P}},-\pmb{\rho},z_{\nu})\,,
\end{eqnarray}
with a ``bubble'' weighted by the form function $\phi_{\lambda}$:

\begin{equation}\label{ghat}
\hat{{\mathcal{G}}}({\mathbf{P}},\pmb{\rho},z_{\nu})\,=\,\sum_k\, e^{i{\mathbf{k}}\cdot\pmb{\rho}}\phi_{\lambda}({\mathbf{k}})\tilde{{\mathcal{G}}}({\mathbf{P}},{\mathbf{k}},z_{\nu})\,.
\end{equation}

The two-fold effect of the interaction potential is clearly visible in this result. On one hand, it leads to a ``renormalized'' one-electron propagator $G$ 
and thus to a bubble ${\mathcal{G}}$ describing the motion of two independent ``renormalized'' quasi-particles. On the other hand, the $T$-matrix represents the scattering processes that finally lead to the formation of the preformed  
pairs.  Following expression (\ref{defdistr}) we can thus construct a function, $g^0_P$, that represents  the internal structure of two independent particles with center of mass momentum ${\mathbf{P}}$ and which is given by the first term in (\ref{chiTm}):
\begin{eqnarray}
g_P^0({\mathbf{P}},\pmb{\rho})&\equiv &\int \,d\Omega\, N_B(\Omega)\, Im{\{\chi^0({\mathbf{P}},\pmb{\rho},\pmb{\rho},z_{\nu})|_{z_{\nu} \to \Omega+i 0^+}\}}\nonumber \\
&\equiv &\int \,d\Omega\, N_B(\Omega)\,Im{\{{\mathcal{G}}({\mathbf{P}},{\mathbf{0}},z_{\nu})|_{z_{\nu} \to \Omega+i  0^+}\}}\,.
\end{eqnarray}
Being independent of $\pmb{\rho}$, this contribution forms a constant ``background'' 
to the structure of $g_P$, resulting from the direct two-particles scattering. For this reason, we subtract this background contribution from 
$g_P$, so that instead of the defintion (\ref{defPhi}) we will use the "rescaled":
\begin{equation}\label{defPhi'}
\Phi({\mathbf{P}},\pmb{\rho},\Omega)\,=\,Im{\{\chi({\mathbf{P}},\pmb{\rho},\pmb{\rho},\Omega+i \eta)\,-\,\chi^0({\mathbf{P}},0,\Omega+i \eta)\}}\,.
\end{equation}
However, one has to bear in mind that possible negative values of  $g_P$ simply mean a reduction of the number of corresponding pairs with respect to this independent particle background.

We do not attempt a fully self-consistent calculation of $G$ - and thus of ${\mathcal{G}}$ - and $\chi$. Rather, we approximate the one-electron spectral function in the 
pseudogap regime we are interested in by a BCS-like form with a finite linewidth $\gamma$. The corresponding result for ${\mathcal{G}}$ is derived in Appendix A. 

In the following section we will discuss the general behavior and present numerical results for the $T-$matrix and the pair structure function for the $s-$ and $d-$wave pairing symmetries.

\section{Numerial results}\label{sec:3}

Our numerical calculations are based on the tight-binding spectrum resulting 
from the kinetic energy part of the Hamiltonian in (\ref{hamilt}) with a nearest neighbor hopping $t$ in two dimensions. We use the total bandwidth $W\,=\,8t$ as our energy 
unit. The zero of the energy is in the centre of the band. Assuming that the underdoped cuprates with a pronounced pseudogap are situated in an intermediate coupling regime of the attractive Hubbard model \cite{sewer} we chose an interaction strength $U$ is of the order of $0.5\,W$. Numerical simulations 
\cite{singer} of the attractive Hubbard model with a local interaction of this 
strength in two dimensions point to a critical temperature $T_c$ on the order 
of $0.01\,W$ \emph{for s-wave symmetry}. 
This result has to be taken as a qualitative indication, since it is not clear 
 whether quantum Monte Carlo calculations can really represent correctly the 
Kosterlitz-Thouless-type (KT) phase transition that is expected in two dimensions. From Quantum Monte Carlo calculations, the gap value in the s-wave case 
for the interaction strength half of the bandwidth, is of the order of $t/2$. 

Based on these numerical results from simulations, we choose a gap parameter value for which the
 critical temperature is approximately $0.2 \Delta$ in the $s-$wave case. In the  $d-$wave case, however, the interaction strength should be of order of the bandwidth to obtain reasonable values for the critical temperature,\cite{Engelbercht} and still maintain the proportion between  the critical temperature and the value of the pseudogap  approximately equal to $0.2$. 

In order to incorporate the effect of the finite lifetime of the pair excitations without washing out  the effect of the pseudogap, we choose the finite linewidth $\gamma$ (see Appendix A) to be of the order of $\Delta/2$. In this way, in fact, one can 
still see rather sharp features in the pseudogap region, while for bigger values of $\gamma$ they are strongly reduced.

The relevant temperature interval of the pseudogap regime is determined in the 
framework of our description of pairing. In the usual T-matrix approximation, 
which results from the Bethe-Salpeter equation (\ref{BS}), the superconducting 
instability occurs at the temperature where the ``Thouless criterion'' is fulfilled:
\begin{equation}\label{Thcr}
T^{-1}({\mathbf{P}}\,=\,0,\,z_{\nu}\,=\,0)\,=\,0\,,
\end{equation}
i.e., where pairs with zero centre of mass momentum begin to form a true bound 
state. This is a ``mean-field-like'' criterion that corresponds to our approach
 to pair structure, which is, of course, less precise than KT, but the latter would require to go more deeply into the analysis of phase fluctuations.

\begin{figure}
\centering{
\includegraphics[height=8.3cm,width=4.2cm,angle=-90]{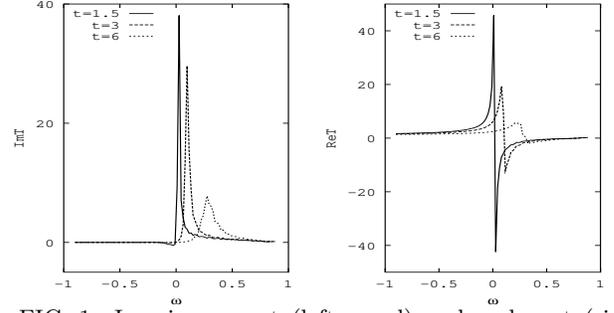}
}
\caption{Imaginary part (left panel) and real part (right panel) of the T-matrix for the $d-$wave case for different temperatures $t=T/T_c$ and ${\mathbf{P}} = 0$. }
\label{fig:01}
\end{figure}

In Figure \ref{fig:01} we show the effect of temperature on the real and the imaginary parts of the T-matrix $T({\mathbf{0}},\Omega +i 0^+)$  for the case ${\mathbf{P}} = 0$. As the temperature approaches the transition value, the peaks in the real and imaginary part of the $T-$ matrix become sharper, and move toward $\omega = 0$. The inverse of the width of the imaginary part peak can be interpreted as the lifetime of the preformed pairs. 
We obtain for the lifetime the values $25,\,  5.5,$ and  $1.8$, for $T/T_C$ equals to $1.5$,  $3$ and $6$, respectively. The lifetime is expressed in unit of the inverse of the Fermi energy $E_F$, where $E_F=\mu-E_{bottom}$ and $E_{bottom}$ is the bottom of the band energy value.

\noindent
For our calculation, we will then consider the two temperature ratio: $T/T_c=1.5$ and $T/T_c=3$ to constrast the cases in which the quasiparticle features are more or less pronounced.

 We study the behaviour of the function $g_P$ for variuos values of the pair 
internal momentum. Moreover, we also investigate the influence of temperature,
 chemical potential and interaction strength on the shape of $g_P$.

For comparison, we will also show the results for $\Delta = 0$ for $s-$ and  $d-$wave interaction symmetries. Although the pair structure function is defined on a lattice, we present it as
 a continuous function, as a guideline for the eye
 to follow its dependence on the 
relative distance.

\subsection{$s-$wave pairing}

\begin{figure}
\centering{
\includegraphics[width=8.6cm,height=8.8cm]{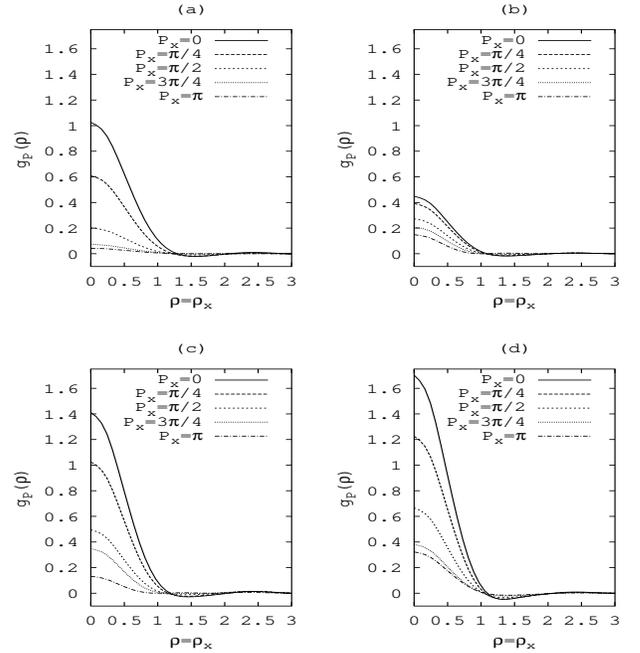}
}
\caption{``Pair structure function'' $g_P(\pmb{\rho})$ for $\pmb{\rho}=(\rho,0)$ (in units of $a$)  and $\Delta = 0$ and:
 (a) $T = 1.5 T_c$, $\mu = -0.4 W$, $U = 0.5 W$; (b) $T = 3 T_c$, $\mu = -0.4 W$, $U = 0.5 W$; (c) $T = 1.5 T_c$, $\mu = -0.3W$, $U = 0.5 W$; (d) $T = 1.5 T_c$, $\mu = -0.4 W$, $U = 0.6 W$.}
\label{fig:1}
\end{figure}

We begin by considering the simplest case of $\Delta=0$, in order to show the general tendence of the pair-structure function without the complication of the influence of the pseudogap on the pairs.
In Figure \ref{fig:1}, we show a typical set of curves for  $s-$wave symmetry pairing for the case $\Delta=0$. The function is calculated for the relative distance  directed along the $x$ axis in the Brillouin zone. In the $s-$wave case, however, the pair structure function is isotropic, reflecting the spatial symmetry of the order parameter, so that $g_P({\mathbf{P}},\pmb{\rho}_x)$ is representative of the behaviour along the other directions.  
Moreover, the pair structure function depends mainly on the modulus of the centre of mass momentum, while the variations of the direction of $\mathbf{P}$ have no relevance for $g_P$.
As can be easily seen, the effect of increasing pair internal momentum is to reduce the number of 
pairs and to shrink their size (which can be seen as a sort of ``relativistic length contraction''). It is  apparent that the extension of the pairs in all the cases here considered is of the order of the lattice constant.
The effect of the temperature on the distribution function is seen by comparing  Figure  \ref{fig:1}a and  \ref{fig:1}b. As expected, raising the 
temperature reduces the number of pairs. A slight decrease of the avarage extension 
of the pair can also be observed. We notice also that increasing  the temperature tends to 
wash out the difference of population of pairs with different internal momentum. In fact, due to the increase of 
the kinetic energy for particle, it becames easier to form pairs with finite center of mass momentum, so that the number of the pairs with ${\mathbf{P}} \neq 0$ increases, even if the net effect of increasing temperature is to lower the total number of pairs.

In Figure  \ref{fig:1}c we show $g_P$ for a different value of the chemical potential 
$\mu$. The comparison is made with Figure \ref{fig:1}a, at a fix temperature ratio with respect to the critical one: $T/T_c = 1.5 $. As the chemical potential moves up from the bottom of the band (i.e. for 
increasing density, but still in the limit of validity of our $T-$matrix approximation), the number of pairs increases, while the size of the pair slightly decreases.
Fixing the chemical potential and for a stronger interaction, we can observe that  the number of pairs increases,
 while their spatial extension does not vary appreciably (Figure  \ref{fig:1}d).

\begin{figure}
\centering{
\includegraphics[width=8.6cm,height=8.8cm]{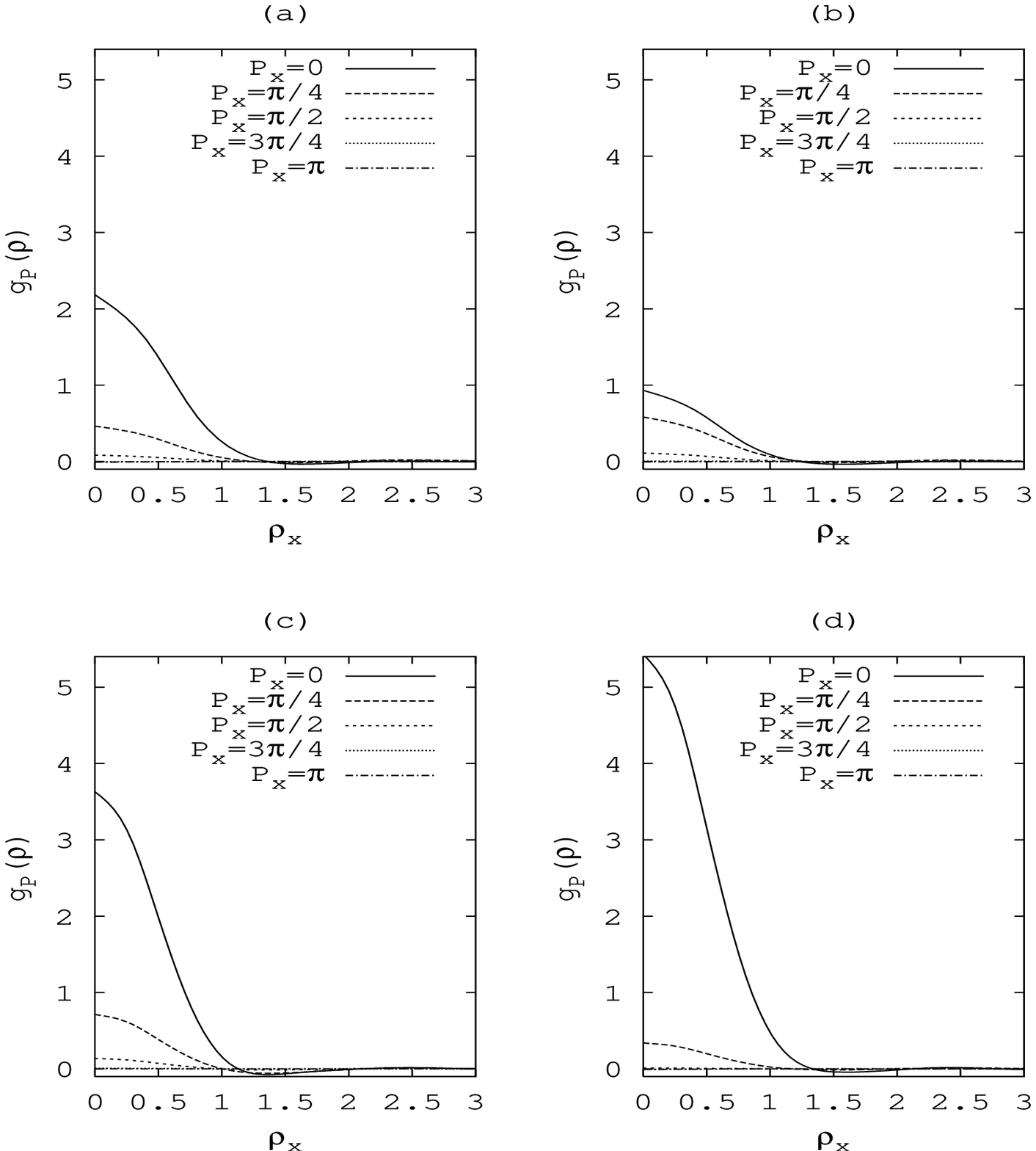}
}
\caption{``Pair structure function'' $g_P(\pmb{\rho})$ for:
 (a) $\Delta = 0.04 W$, $T = 1.5 T_c$, $\mu = -0.4 W$, $U = 0.5 W$; (b) $\Delta = 0.04 W$, $T = 3 T_c$, $\mu = -0.4 W$, $U = 0.5 W$; (c) $\Delta = 0.062 W$,  $T = 1.5 T_c$, $\mu = -0.3W$, $U = 0.5 W$; (d) $\Delta = 0.067 W$, $T = 1.5 T_c$, $\mu = -0.4 W$, $U = 0.6 W$.}
\label{fig:3}
\end{figure}

We present now the results for the pseudogap region, i.e. for $\Delta \neq 0$ (Figure \ref{fig:3}). The effect of a pseudogap opening in the single particle spectrum 
is to increase the number of pairs and to enlarge their size for zero internal momentum, while for finite value of ${\mathbf{P}}$ it seems that the opening of a pseudogap suppress the pairing, as one can see comparing Figures \ref{fig:1}a and \ref{fig:3}a. 
As already mentioned, we choose the parameters for the pseudogap region from the results of Monte Carlo 
simulation.\cite{sewer} The dependence of the (pseudo)gap parameter $\Delta$ 
on the chemical potential and on the interaction  has been stated by fixing the relative magnitude of the critical temperature and the pseudogap amplitude to be the one tipical for cuprates.

\begin{figure}
\centering{
\includegraphics[width=4.5cm,angle=-90]{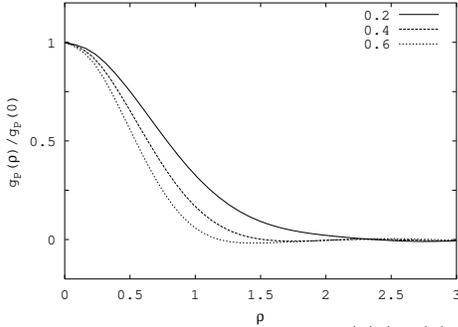}
}
\caption{``Pair structure function'' $g_P(\pmb{\rho})/g_P(0)$ for $\Delta = 0$ and $U\,=\, 0.2W$ (full line), $0.4W$ (dashed) and $0.6W$ (dotted).}
\label{fig:4}
\end{figure}

As we noticed, the opening of a pseudogap strongly depresses the number of 
pairs having a finite momentum. This behaviour becomes even more evident when 
 the strength of the interaction increases (Figure \ref{fig:3}d), which in turn
 implies a bigger value of the pseudogap. A part from the different behaviour at finite momentum, the qualitative dependence of the 
pair distribution function on the parameters in presence of a pseudogap does 
not differ  from the case $\Delta = 0$. Also, we have not found any substantial variation of the avarage internal distance of the pairs by varying the interaction strength. This is in part due to the fact that we confine our study to the pseudogap region, so that the range of variation for the interaction strength is not sufficient to crossover from Bose-Einstein condensation to BCS behaviour. In Figure (\ref{fig:4}) we show the variation of the pair internal distance with respect to variation of the interaction strength for the case $\Delta\, =\, 0$, where we have normalized the pair structure function to its zero internal distance value, as we want to emphasize the trend of the functions. As it can be easily seen, the internal distance has doubled as $U$ has passed from $0.6 W$ to $0.2 W$. 

\noindent
This evolution is similar to what has been obtained by Paredes and Cirac \cite{para} for a system of bosonic atoms in an optical lattice, which is mapped into a fermionic system via a Jordan-Wigner transformation with a local attraction $V$. The pairs are formed according to a variational wave function, whose spatial extension changes by changing the attraction strength, and results in localized pairs for large value of $V$, and delocalized pairs for small $V$.

\begin{figure}
\centering{
\includegraphics[width=8.6cm,height=8.4cm,angle=-90]{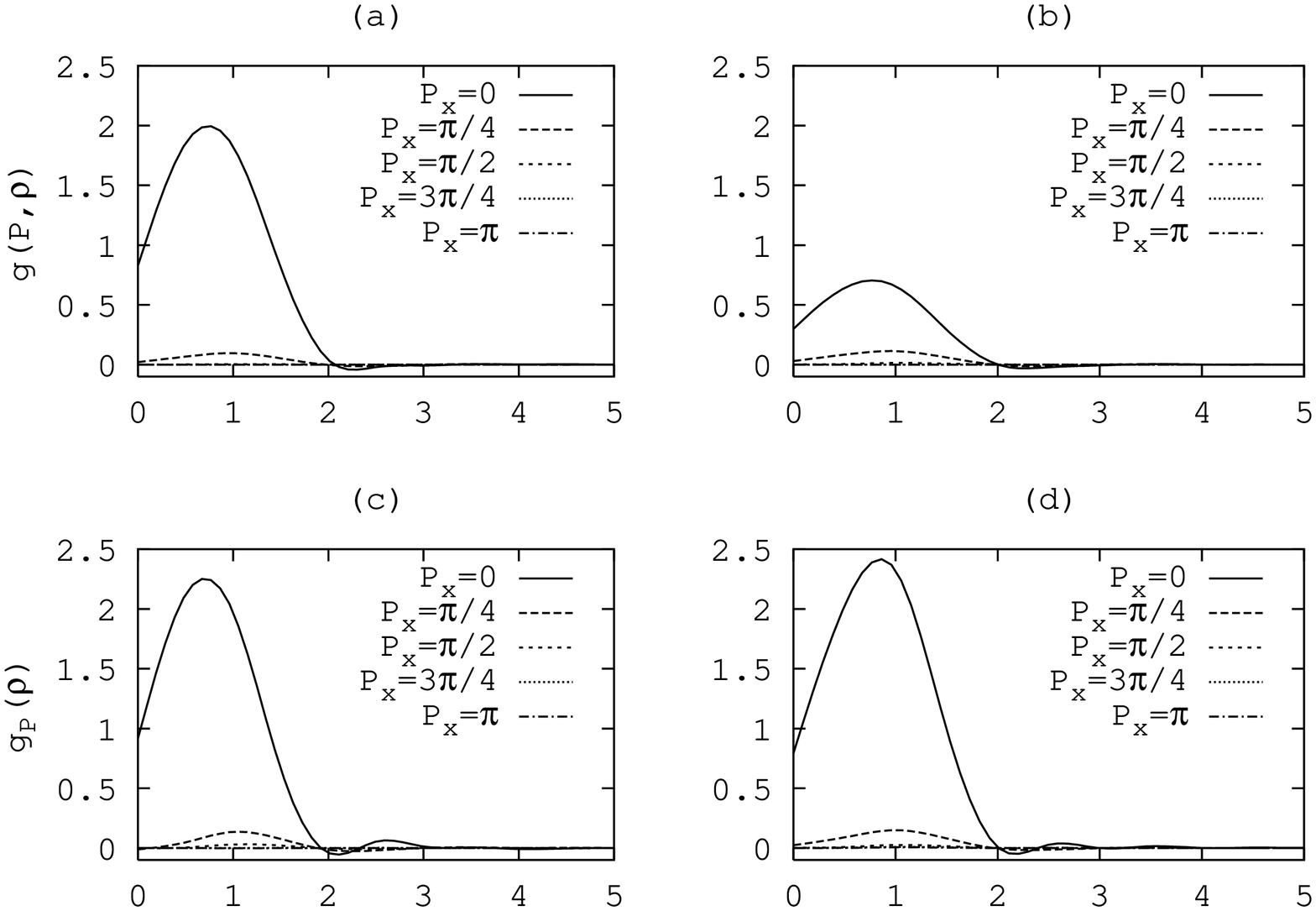}
}
\caption{``Pair structure function'' $g_P(\pmb{\rho})$ for:
 (a) $\Delta = 0.047 W$, $T = 1.5 T_c$, $\mu = -0.4 W$, $U = 0.5 W$; (b) $\Delta = 0.047 W$, $T = 3 T_c$, $\mu = -0.4 W$, $U = 0.5 W$; (c) $\Delta = 0.054 W$,  $T = 1.5 T_c$, $\mu = -0.3W$, $U = 0.5 W$; (d) $\Delta = 0.078 W$, $T = 1.5 T_c$, $\mu = -0.4 W$, $U = 0.7 W$.}
\label{fig:4.1}
\end{figure}

In the context of cuprate supercondutors, it is interesting also to consider the extended $s-$wave symmetry, which has been recently proposed to explain some experiments.\cite{zhao} In Figure \ref{fig:4.1} we present the pair structure function calculated in the pseudogap region for $\phi_{\lambda }({\mathbf{k}})=\cos{k_x}+\cos{k_y}$.
We can see that in this case the maximum of the distribution is shifted from zero internal distance to a finite distance of the order of one lattice constant. However, the effects of the variation of the parameters on the pair structure function are the same as in the full gapped $s-$wave.

\subsection{$d-$wave pairing}

\begin{figure}
\centering{
\includegraphics[width=8.6cm,height=8.8cm]{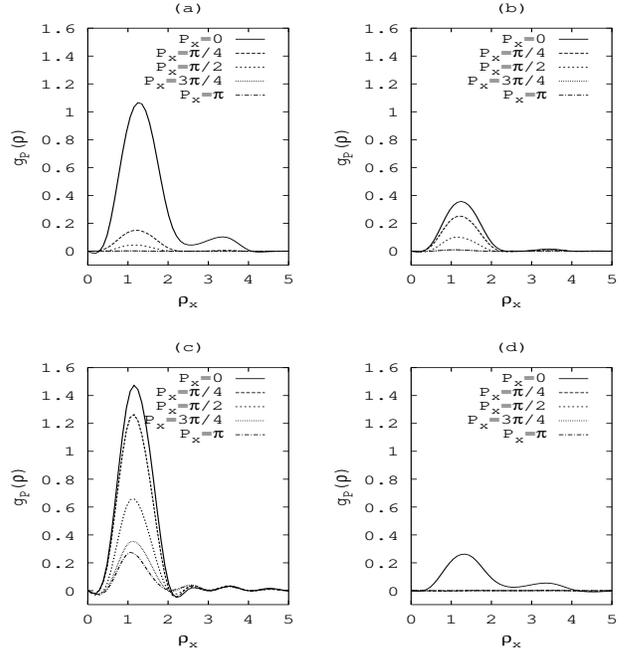}
}
\caption{``Pair structure function''  $g_P(\pmb{\rho})$  for $d-$wave case and $\Delta = 0$: (a) $T = 1.5 T_c$, $U = 0.65 W$, $\mu = -0.4 W$; (b) $T = 3 T_c$, $U = 0.65 W$, $\mu = -0.4 W$; (c) $T = 1.5 T_c$, $U = 0.65 W$, $\mu = -0.3 W$; (d) $T = 1.5 T_c$, $U = 0.63 W$, $\mu = -0.4 W$.}
\label{fig:5}
\end{figure}

The $d-$wave case does not differ too much in the qualitative behaviour from the previous cases. Nevertheless, in the $d-$wave case, $g_P$ vanishes for zero internal distance, as it should be from the symmetry of the interaction and of the pseudogap parameter, and has its maximum at a distance of the order of a lattice constant, slightly higher than the extended $s-$wave case. The fact that the pairs  have again typically the size of a lattice constant is a consequence of our nearest neighbor type of attraction. 
As for the $s-$wave case, we first analize the case $\Delta=0$ (Figure \ref{fig:5}).  Besides the general features alreaady mentioned, we can see that a secondary peak appears in the zero momentum distribution function at a distance of three or four lattice constant.  Moving pairs show a slight contraction and they are less numerous than pairs at rest. 
As the temperature is raised, the distribution function is suppressed for $P=0$ and enhanced for $P \neq 0$, as in the $s-$wave case (compare Figure \ref{fig:5}a and \ref{fig:5}b). 

\noindent
By lowering the interaction strength (Figure \ref{fig:5}a and \ref{fig:5}d), the number of pairs with internal momentum ${\mathbf{P}}$ tends to diminish strongly. Moving the chemical potential more inside the band (Figure \ref{fig:5}a and \ref{fig:5}c) has as a result an enhancement of the number of pairs. In this case, the increase in the population of pairs with finite momentum is remarkable, in contrast to the behaviour obtained when the interaction strength is changed, when no noticeable change in the relative populations for different ${\mathbf{P}}$ appears. The result obtained by varying the chemical potential can be attributed to a greater avarage kinetic energy due to the spatial confinement of the particle at higher densities.
The form of $g_P$ shows the typical Friedel oscillations, the wave length of which depends on electronic 
density (as can be seen when changing the value of $\mu$). 

\begin{figure}
\centering{
\includegraphics[width=8.6cm,height=8.8cm]{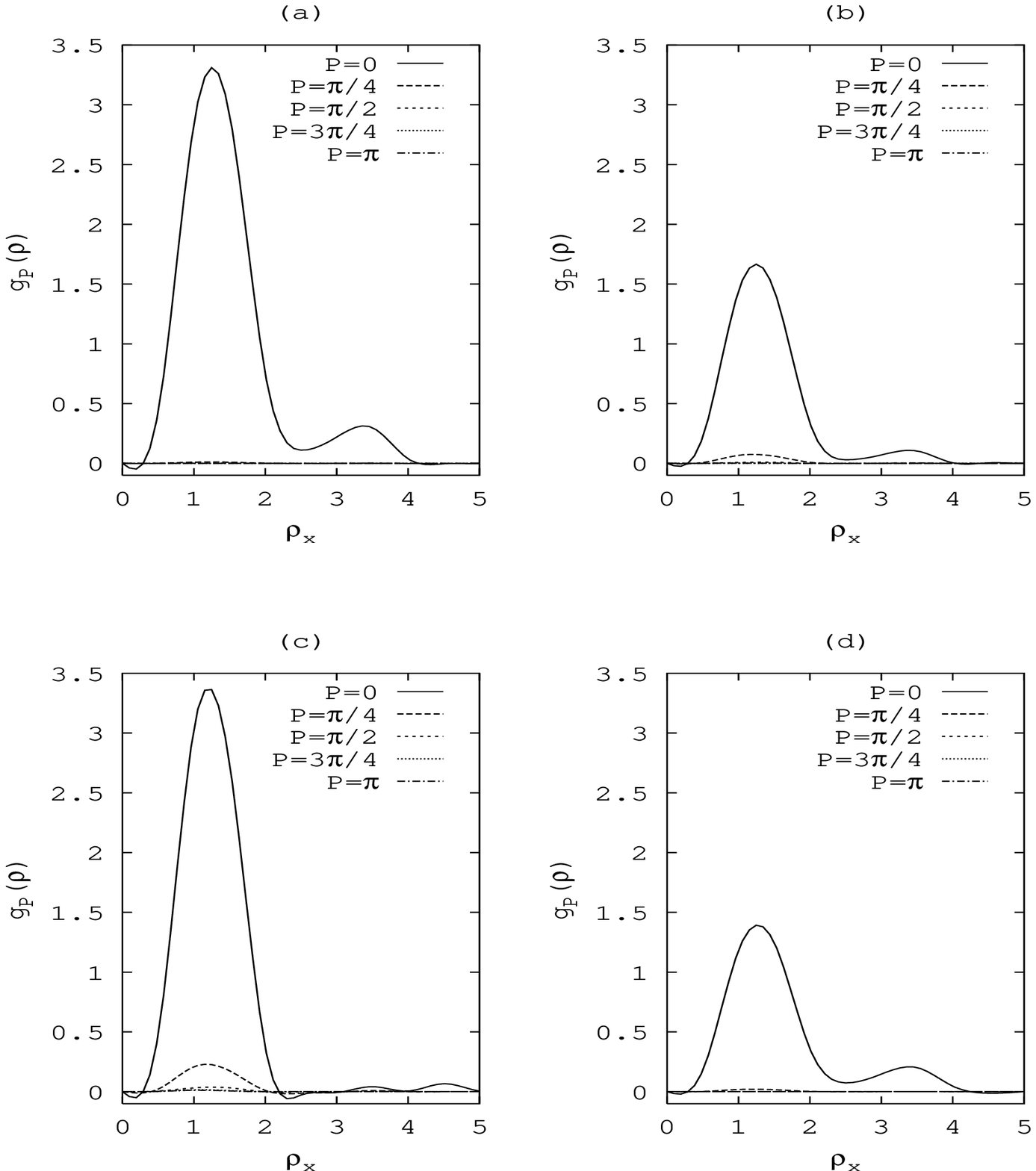}
}
\caption{``Pair structure function'' $g_P(\pmb{\rho})$ for:
 (a) $\Delta = 0.05 W$, $T = 1.5 T_c$, $\mu = -0.4 W$, $U = 0.7 W$; (b) $\Delta = 0.05 W$, $T = 3 T_c$, $\mu = -0.4 W$, $U = 0.7 W$; (c) $\Delta = 0.1 W$,  $T = 1.5 T_c$, $\mu = -0.3W$, $U = 0.7 W$; (d) $\Delta = 0.04 W$, $T = 1.5 T_c$, $\mu = -0.4 W$, $U = 0.68 W$.}
\label{fig:6}
\end{figure}

The opening of a pseudogap (Figure \ref{fig:6}a) strongly enhances the number of pairs for total momentum $P=0$, while suppresses the number of pairs with $P \neq 0$. No change in the pair size is instead noticeable. Another fact to be noticed is that when a gap opens we need to move toward stronger interaction value in order to have a reasonable value for the transition temperature. This is due to the fact that a finite value of the interaction in the 2D Hubbard model is needed for the $d-$wave case to have a bound state in the related two-body problem.\cite{APPS} Moreover, the Friedel oscillations are in this case hidden by the oscillation of the $d-$wave symmetry of pseudogap parameter, of typical period $\pi /a$.

\begin{figure}
\centering{
\includegraphics[width=6.8cm,angle=-90]{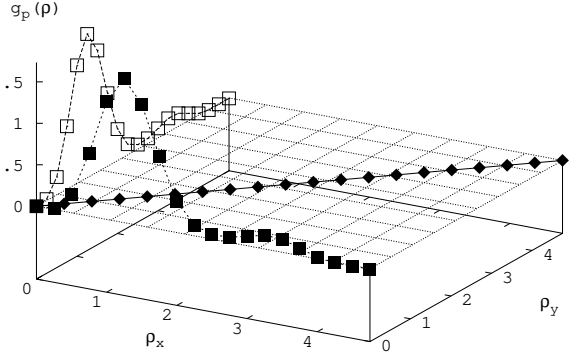}
}
\caption{Spatial structure of ``pair structure function'' $g_P(\mathbf{0},\pmb{\rho})$ for the $d-$wave symmetry ($\Delta = 0.05 W$, $T = 1.5 T_c$, $\mu = -0.4 W$, $U = 0.7 W$)}
\label{fig:7}
\end{figure}

As expected, we obtained that the $d$-wave pairs are 
elongated in the direction of the coordinate axis, whereas their width shrinks to zero for $\pmb{\rho}$ along the diagonal ($\rho_x = \rho_y$) (see Figure \ref{fig:7}). As in the $s-$wave case, the pair structure function is not very sensitive to the direction of $\mathbf{P}$, but depends only on its magnitude.

The effects of temperature and interaction strength on the pair structure function are not different in the general behaviour from the case with no pseudogap.
To increase the chemical potential  depresses strongly the zero momentum pair distribution. On the contrary, the populations of pairs with finite momentum increase. To explain this  enhancement, we can invoke the same explanation as in the $\Delta = 0$ case (namely, the increase of kinetic energy due to confinement). We can also note a decrease of the avarage extension of the pair, as the secondary peak in the pair structure function in figure \ref{fig:6}c has almost completely disappeared. It is to note, however, that the total number of pairs seems to decrease with increasing chemical momentum, despite the fact that the total number of particles is higher. This could  be attributed to a breakdown of our $T-$matrix approximation, or to the fact that we are approaching a region in which the pseudogap is closing.

\section{Discussion}\label{sec:4}
In this work, we have investigated the internal structure of preformed pairs in the pseudogap region of high temperature superconductors. We assume that this region can be described by an attractive Hubbard model. In this framework, we have studied the Green function for the operator which creates a pair 
of particles whose partners are at a certain relative distance. We applied the  Bethe-Salpeter approach and introduce a $T-$matrix formalism, which takes into account the relevant interaction channel of repeated two-particle scattering.
 Our calculations show that in a system described by an attractive Hubbard 
model, the opening of a pseudogap in the one particle spectrum increases the 
probability of finding two particles at a distance of the order of the lattice 
spacing, with respect to the case in which no gap is present. We have also found that in the range of values that we have considered for the parameters (and that should correspond to the pseudogap region) the average extension of the pair is of the order of a lattice constant. The interparticle distance does not change appreciably by 
varying the interaction strength, but this is not too surprising since we have not investigated the BCS limit of our model, where the pseudogap region is believed not to be present. Nevertheless, in the broken symmetry phase the dependence of $\xi_{pair}$ at $T = 0$ on the interaction strength is instead quite strong\cite{{PistolesiStrinati},{Benfatto}}, such that one could presume that the anomalous Green functions might in fact bear the major effect to the extension of the pair.
The effect of increasing temperature is to reduce the total number of preformed pairs and to shorten their internal distance. Also, by increasing the avarage kinetic energy of the particle, it changes the relative weight of the populations corresponding to different centre of mass momentum  ${\mathbf{P}}$, depressing the one with ${\mathbf{P}}=0$ more than the ones with ${\mathbf{P}} \neq 0$.
The influence of the chemical potential on the pair distribution function reflects the influence of the density of particles. It is not surprising, then, that moving the chemical potential more inside the band (i.e., for higher densities) increases the number of pairs with higher momentum: the kinetic energy of the particle should in fact increase as well, and so does the avarage momentum of the pairs, since the number of pairs for a given momentum is given by the integral over the internal distance of the pair structure function (see Appendix B).

One interesting question concerns the connection between the preformed pairs and the structure of Cooper pairs below $T_c$. We in fact expect to recover in the broken symmetry phase the usual BCS to Bose-Einstein condensation behaviour, which obviously affects  also the average internal pair distance.

Another interesting observation related to the pair correlation function, which is equivalent to our pair structure function integrated over all the centre of mass momenta, is the fact that  in the broken symmetry phase it is connected to the correlation energy.\cite{vandermarel} This energy, in turn, plays a key role in the thermodynamics of the phase transition itself. The real space picture of this  pair correlation function, reported in \cite{vandermarel} looks similar to what we obtained for $g_P(P,\pmb{\rho})$ above $T_c$ both for $s-$ and $d-$wave case as can be seen by comparing with Figure \ref{fig:7} (we point out that in contrast to reference \cite{vandermarel}, we reported the values of the pair structure function not only on the lattice sites, but also on intermediate positions).

Besides this aspect related to superconductors, another possible relevance for the internal structure of Cooper pairs is in the very active field of systems of ultracold atoms. In fact, while it is very difficult to measure the internal structure of the Cooper pairs in a superconductor, due to the very complicated structure of the cuprates, in the case of atoms trapped in optical lattice it is possible to carefully control all the parameters for the interaction among particles\cite{GreinerOrzel}. Moreover, it should be possible to actually measure the extension of the Cooper pairs, for example, by analyzing the response to a Raman laser which couples to the internal states of the pair, as suggested in \cite{para}.

In order to make a more meaningful comparison with the cited papers dealing with the superconducting phase, we should however calculate the pair structure function below $T_c$. That will be the subject of a successive work.

We thank S. Sharapov, P. Pieri and G. C. Strinati for helpful discussions. This work was supported by the National Competence Centre "MaNEP" of the Swiss National Science Foundation.

\appendix

\section{Spectral function in the pseudogap region}\label{app:A}

To mimic the presence of a pseudogap, we consider the form of the BCS Green function with a finite linewidth. Due to the absence of long range  order the ``anomalous'' BCS Green function is  zero. We have then:
\begin{equation}\label{gbcs1}
G({\mathbf{k}},z)\,=\,\frac{1}{z-\xi({\mathbf{k}})+i\Gamma-\frac{\Delta^2({\mathbf{k}})}{z-\xi({\mathbf{k}})+i\Gamma}}\,,
\end{equation}
where $\xi({\mathbf{k}})\,=\,-2\,t(\cos{k_x}+\cos{k_y})$ is the single particle disperion, $1/\Gamma$ is the excitation lifetime and $\Delta({\mathbf{k}})$ is the pseudogap.
It is possible to rewrite (\ref{gbcs1}) in terms of the quasiparticles excitation energy, that is  $E({\mathbf{k}}) = \sqrt{\xi^2({\mathbf{k}}+\Delta^2({\mathbf{k}})}$:
\begin{eqnarray}\label{gbcs2}
G({\mathbf{k}},z)&=&\frac{z\,+\,\xi({\mathbf{k}})\,+\,i\Gamma}{(z+E({\mathbf{k}})+i\Gamma)(z-E({\mathbf{k}})+i\Gamma)}\notag \\
&=&\frac{u^2_{{\mathbf{k}}}}{z-E({\mathbf{k}})+i\Gamma}\,+\,\frac{v^2_{{\mathbf{k}}}}{z+E({\mathbf{k}})+i\Gamma}\,,
\end{eqnarray}
where $u^2_{{\mathbf{k}}} = (E({\mathbf{k}})+\xi({\mathbf{k}}))/(2 E({\mathbf{k}}))$ and $v^2_{{\mathbf{k}}} = (E({\mathbf{k}})-\xi({\mathbf{k}}))/(2 E({\mathbf{k}}))$.
From (\ref{gbcs2}) we can deduce the form of the spectral function:
\begin{equation}\label{spectr}
A({\mathbf{k}},z)=\frac{1}{\pi}\left(\frac{u^2_{{\mathbf{k}}}\Gamma}{(z-E({\mathbf{k}}))^2+\Gamma^2}+\frac{v^2_{{\mathbf{k}}}\Gamma}{(z+E({\mathbf{k}}))^2+\Gamma^2}\right)\,.
\end{equation}

The particle-particle bubble can be written in term of the spectral function of the single particle Green functions:
\begin{eqnarray}
{\mathcal{G}}({\mathbf{P}},{\mathbf{k}},z)&=&\sum_{z_{\nu}}\, G({\mathbf{k}}+\frac{{\mathbf{P}}}{2},z_{\nu})\,G({\mathbf{k}}-\frac{{\mathbf{P}}}{2},z-z_{\nu})\notag \\
&=&\int_0^{\infty}\frac{d\omega }{2\pi}\,A({\mathbf{k}}+\frac{{\mathbf{P}}}{2},\omega )\int_0^{\infty}\frac{d\omega '}{2\pi}\,A({\mathbf{k}}-\frac{{\mathbf{P}}}{2},\omega ')\,\notag \\
&\times& \frac{f_F(\omega)+f_F(\omega ')-1}{z-\omega -\omega '}\,.
\end{eqnarray}
Replacing  the expression for the single particle  spectral function (\ref{spectr}) into the previous equation and performing the substitution $z \to \Omega+ i \delta$, we can obtain the expression for the spectral function of the  particle-particle bubble $I({\mathbf{P}},{\mathbf{k}},\Omega)$:
\begin{eqnarray}
&&I({\mathbf{P}},{\mathbf{k}},\Omega)\,\equiv\,Im{{\mathcal{G}}({\mathbf{P}},{\mathbf{k}},\Omega + i \delta)} =\notag \\
&&\frac{2}{\pi}\left(\frac{u^2_1u^2_2\Gamma }{(\Omega-(E_1+E_2))^2+4\Gamma^2}+\frac{v^2_1v^2_2\Gamma }{(\Omega+(E_1+E_2))^2+4\Gamma^2}\right.
\notag \\
&&+\left.\frac{u^2_1v^2_2\Gamma }{(\Omega-(E_1-E_2))^2+4\Gamma^2}+\frac{v^2_1u^2_2\Gamma }{(\Omega+(E_1-E_2))^2+4\Gamma^2}\right)\,,\nonumber \\
&&
\end{eqnarray}
where we used the shorthand notation $1 = {\mathbf{k}}+{\mathbf{P}}/2$ and  $2 = {\mathbf{k}}-{\mathbf{P}}/2$.
The parameter that enters our calculations and that mimics the pseudogap is then the linewidth  $\gamma = 2 \Gamma$ for the excitation described by the particle-particle bubble.

\section{Number of preformed pairs}\label{app:B}

In this Appendix we relate the definition of the pair correlation
 function (\ref{chifunc}) to the number $N_{c.p.}$ of fluctuating Cooper pairs above $T_c$. If the analysis is carried out close enough to the critical temperature, one can define the density of fluctuating Cooper pairs starting from the Ginzburg-Landau (GL) theory\cite{varlamov}. Above $T_c$, the order parameter $\psi(\pmb{\rho})$ has a fluctuating origin and depends on spatial coordinates even in the absence of magnetic field. The fluctuating  thermodynamical potential $\Omega_{(fl)}$ is then:
\begin{eqnarray}
\Omega_{(fl)}\equiv \Omega_{(s)}-\Omega_{(n)} &=&\alpha\int d{{\mathbf{r}}} \left[\epsilon  \mid \psi({{\mathbf{r}}}) \mid^2+\frac{b}{2\alpha}\mid\psi({{\mathbf{r}}})\mid^4 \right. \nonumber \\
&+&\left. \eta_{D}\mid\nabla\psi({{\mathbf{r}}})\mid ^2 \right]\,
\end{eqnarray} 
where $\epsilon = \ln{(T/T_c)} \ll 1$, $\alpha = 1/(4 m \eta_D)$ and $\eta_D$ is the phenomenological GL constant.
 The GL order parameter undergoes equilibrium fluctuations which, close enough to the transition temperature, can be described by a probability distribution\cite{varlamov}:
\begin{equation}\label{prob}
{\cal{P}} \propto \exp{\left[\Omega_{(fl})\right]} = \exp{\left[-\frac{\alpha}{T}\sum_{{\mathbf{k}}}\left(\epsilon\,+\,\eta_D k^2\right)\mid \psi_{{\mathbf{k}}}\mid^2\right]}
\end{equation}
where Fourier transform of the order parameter with respect to the spatial variables has been carried out. The average value for the square of the fluctuating GL order parameter is then:
\begin{equation}
\langle \,\vert \psi_{{\mathbf{k}}}^{(fl)}\vert ^2\,\rangle\,=\,\frac{\int \,\vert \psi_{{\mathbf{k}}}\vert ^2{\cal{P}}({\mathbf{k}})\, d \vert \psi_{{\mathbf{k}}}\vert ^2}{\int\,{\cal{P}}({\mathbf{k}})\, d \vert \psi_{{\mathbf{k}}}\vert ^2}\,.
\end{equation}

The number of fluctuating Cooper pairs $N_{c.p.}$ is determined by the average value of the square of the GL order parameter modulus\cite{Abrikosov}:
\begin{equation}\label{NCP}
N_{c.p.}\,=\,\sum_{{\mathbf{k}}} \langle \,\vert \psi_{{\mathbf{k}}}^{(fl)}\vert ^2\,\rangle\,\exp{i({\mathbf{k}} \cdot {{\mathbf{r}}})}\mid_{ {{\mathbf{r}}} \to 0}.
\end{equation}
The number of fluctuating Cooper pairs has been obtained measuring the Knight shift and the nuclear spin relaxation under magnetic field by Zheng and co-workers\cite{Zheng}. These quantities reflect indeed the pseudogap behaviour, and their  reduction above $T_c$ is strongly magnetic field dependent accordingly to a scaling relation that depends on the effects of the Cooper pair density fluctuations\cite{Zheng}.

To relate the definition  (\ref{NCP}) to our pair correlation
 function (\ref{chifunc}), we consider the partition function for the Hamiltonian (\ref{hamilt}). For the sake of simplicity, we consider the case of on-site interaction $U(i-j)\, =\, U\, \delta_{i,j}$, where $\delta$ is the Kronecker delta function.
By means of a Hubbard-Stratonovic transformation we can rewrite the partition function in term of the classical field $\Delta({\mathbf{P}},\tau)$ and its complex conjugate:
\begin{eqnarray}\label{partf}
Z&=&Tr\, e^{-\beta H}\,=\,Tr\, e^{-\beta H_0}\,{\mathcal{T}} \int {\mathcal{D}}\Delta\, {\mathcal{D}}\Delta^* \exp{}\left[-i\int_0^{-i\beta} \right.\notag \\
&\times & \sum_{{\mathbf{P}}}|U|^{-1}\,|\Delta({\mathbf{P}},\tau)|^2\, +\, \left(Q^{\dagger}_0({\mathbf{P}},\tau)\,\Delta({\mathbf{P}},\tau)+\right .\notag \\
&+&\left.\left. Q_0({\mathbf{P}},\tau)\,\Delta^*({\mathbf{P}},\tau)\right) \right] d\tau\,,
\end{eqnarray}
where $\mathcal{T}$ is the time-ordering operator and $Q^{\dagger}_0({\mathbf{P}},\tau)$ is the Fourier transform with respect to spatial coordinates and time of the local pairing operator defined in (\ref{2ptoper}) for $i\,=\,j$.  $H_0$ is the non interacting part of the Hamiltonian $H$ (Eq. \ref{hamilt}). 
 
In order to obtain the two-particle Green function we are interested in, we introduce the source field, $\eta({\mathbf{P}},\tau)$ and its complex conjugate,  which couples to the pairing operator, to obtain the action term (replacing the exponent of (\ref{partf})):
\begin{eqnarray}\label{freeen}
\tilde{S}&=&\int_0^{-i\beta}\,\sum_{{\mathbf{P}}}\,|U|^{-1}\,|\Delta({\mathbf{P}},\tau)|^2 +Q^{\dagger}_0({\mathbf{P}},\tau)\left(\Delta({\mathbf{P}},\tau)+\right.  \notag \\
&+&\left.\eta({\mathbf{P}},\tau)\right)+Q_0({\mathbf{P}},\tau)\left(\Delta^*({\mathbf{P}},\tau)+\eta^*({\mathbf{P}},\tau)\right)\,d\tau\,.
\end{eqnarray}
If we now differenciate the partition function $Z$ with respect to the fields $\eta$ and $\eta^*$, we obtain the Green function for the local pairing operator:
\begin{eqnarray}
\frac{\delta}{\delta \eta^*({\mathbf{P}}',\tau')}\frac{\delta}{\delta \eta({\mathbf{P}},\tau)}\log{Z}\mid_{\eta=0,\eta^*=0}\,&&=\nonumber \\
 =\,\langle T\left[ Q^{\dagger}_0({\mathbf{P}},\tau)\right. &&\left.Q_0({\mathbf{P}}',\tau')\right]\rangle\,.
\end{eqnarray}

We can now define a new field: $\xi({\mathbf{P}},\tau)\equiv \Delta({\mathbf{P}},\tau) + \eta({\mathbf{P}},\tau)$, so that Eq. (\ref{freeen}) becomes:
\begin{eqnarray}\label{freeen1}
\tilde{S}_1&=&\int_0^{-i\beta}\,\sum_{{\mathbf{P}}}\,|U|^{-1}\,|\xi({\mathbf{P}},\tau)-\eta({\mathbf{P}},\tau)|^2 + \notag \\
&+& Q^{\dagger}_0({\mathbf{P}},\tau)\,\xi({\mathbf{P}},\tau)+Q_0({\mathbf{P}},\tau)\xi^*({\mathbf{P}},\tau)\,d\tau\,.
\end{eqnarray}
Differentiating $Z$ again with respect to $\eta$ and $\eta^*$ we obtain the equality:
\begin{eqnarray}\label{relchidel}
\langle T\left[ Q^{\dagger}_0({\mathbf{P}},\tau)Q_0({\mathbf{P}}',\tau')\right]\rangle\mid_{\tau'=\tau}&=&\frac{\langle \Delta^*({\mathbf{P}},\tau)\Delta({\mathbf{P}}',\tau)\rangle}{U^2}\nonumber \\
&-&\frac{i}{|U|}\,,
\end{eqnarray}
\noindent
where we have set $\tau'=\tau$ in order to discuss the relation among this expression and the equation for the number of Cooper pairs derived from the (time independent) GL theory. 
 Above the critical temperature, the fluctuating GL order parameter is proportional to the classical field $\Delta({\mathbf{P}},\tau)$ we have introduced. Equation (\ref{relchidel}) essentially confirms formula (\ref{NCP}), which represents the number of fluctuating Cooper pairs that have \emph{a fixed distance} between the components, namely $\pmb{\rho} = 0$ for an on-site interaction.

We want now calculate the Green function for the $Q$ operator for an internal distance $\pmb{\rho}$. To this end, we must insert in (\ref{partf}) a source field that depends on $\pmb{\rho}$, namely $\eta({\mathbf{P}},\pmb{\rho},\tau)$:
\begin{eqnarray}\label{freeenrho}
\tilde{S}&=&\int_0^{-i\beta}\,\sum_{{\mathbf{P}}}\,|U|^{-1}\,|\Delta({\mathbf{P}},\tau)|^2 +Q^{\dagger}_0({\mathbf{P}},\tau)\Delta({\mathbf{P}},\tau) \notag \\
&+& Q_0({\mathbf{P}},\tau)\Delta^*({\mathbf{P}},\tau) \sum_{\pmb{\rho}}\left(Q^{\dagger}({\mathbf{P}},\pmb{\rho},\tau)\eta({\mathbf{P}},\pmb{\rho},\tau)\right. \nonumber \\
&+&\left.Q({\mathbf{P}},\pmb{\rho},\tau)\eta^*({\mathbf{P}},\pmb{\rho},\tau)\right)\,d\tau\,.
\end{eqnarray}
The correlation function $\chi({\mathbf{P}},\pmb{\rho},\pmb{\rho}',t)$ can be obtained by differentiating  with respect to the source fields:
\begin{eqnarray}\label{chieta}
\frac{\delta}{\delta \eta^*({\mathbf{P}}',\pmb{\rho}',\tau')}\frac{\delta}{\delta \eta({\mathbf{P}},\pmb{\rho},\tau)}\log{Z}&&\mid_{\eta=0,\eta^*=0}\,=\nonumber \\
 =\,\langle T\left[ Q^{\dagger}({\mathbf{P}},\pmb{\rho},\tau)\right. &&\left.Q({\mathbf{P}}',\pmb{\rho}',\tau')\right]\rangle\,.
\end{eqnarray}
\noindent
In order to obtain the relation between (\ref{chieta}) and the classical field $\Delta$, we perform a second order cumulant expansion for the trace over electronic degrees of fredoom in $Z$. 
The odd power terms of the expansion vanish. We then obtain:
\begin{eqnarray}\label{cumul}
Z&=& Z_0 \int {\mathcal{D}}\Delta {\mathcal{D}}\Delta^* \exp{\left[-i\int_0^{-i\beta}\,\sum_{{\mathbf{P}}}|U|^{-1}|\Delta({\mathbf{P}},\tau)|^2 d\tau \right]} \notag \\
&\times &\exp{\left[-\frac{1}{2}\int \int\,\sum_{{\mathbf{P}},\,{\mathbf{P}}'}C({\mathbf{P}},\tau;{\mathbf{P}}',\tau') d\tau d\tau'\right]}\,,
\end{eqnarray}
where:
\begin{eqnarray}
&&C({\mathbf{P}},\tau;{\mathbf{P}}',\tau')=\nonumber \\
&&\sum_{\pmb{\rho},\pmb{\rho}'}\langle T\left[Q^{\dagger}({\mathbf{P}},\pmb{\rho},\tau)Q({\mathbf{P}} ',\pmb{\rho} ',\tau ')\right]\rangle_c \, \eta({\mathbf{P}},\pmb{\rho},\tau)\eta^*({\mathbf{P}}',\pmb{\rho}',\tau')\,+ \notag \\
&&+\sum_{\pmb{\rho}}\langle T\left[ Q^{\dagger}({\mathbf{P}},\pmb{\rho},\tau)Q_0({\mathbf{P}} ',\tau ')\right]\rangle_c  \, \eta({\mathbf{P}},\pmb{\rho},\tau)\Delta^*({\mathbf{P}}',\tau')\,+ \notag \\
&&+\sum_{\pmb{\rho}'}\langle T\left[Q^{\dagger}_0({\mathbf{P}},\tau)Q({\mathbf{P}} ',\pmb{\rho} ',\tau ')\right]\rangle_c  \, \Delta({\mathbf{P}},\tau)\eta^*({\mathbf{P}}',\pmb{\rho}',\tau')\,+ \notag \\
&&+\langle T\left[Q^{\dagger}_0({\mathbf{P}},\tau)Q_0({\mathbf{P}} ',\tau ')\right]\rangle_c  \, \Delta({\mathbf{P}},\tau)\Delta^*({\mathbf{P}}',\tau')\,+\,h.c.\,.\notag \\
&&
\end{eqnarray}
$\langle \cdots \rangle_c$ is the second order cumulant, which in our case coincides with the avarage over $H_0: \, \langle \cdots \rangle_0$.
Applying the definition (\ref{chieta}) to the expression (\ref{cumul}), we obtain:
\begin{eqnarray}\label{chicum}
&&\langle T\left[ Q^{\dagger}({\mathbf{P}},\pmb{\rho},\tau)\,Q({\mathbf{P}}',\pmb{\rho}',\tau')\right]\rangle_2 \,=\nonumber \\
&&-\langle T\left[ Q^{\dagger}({\mathbf{P}},\pmb{\rho},\tau)Q({\mathbf{P}}',\pmb{\rho}',\tau')\right]\rangle_0 + \int \int d\tau'' d\tau''' \sum_{{\mathbf{P}''},{\mathbf{P}'''}} \notag \\
&&\times \langle T\left[ Q^{\dagger}({\mathbf{P}}'',0,\tau'')Q({\mathbf{P}}',\pmb{\rho}',\tau')\right]\rangle_0 \langle T\left[Q^{\dagger}({\mathbf{P}},\pmb{\rho},\tau)\right. \nonumber\\
&&\times \left.Q({\mathbf{P}}''',0,\tau''')\,\right]\rangle_0 \langle\, \Delta({\mathbf{P}}'',\tau'') \Delta^*({\mathbf{P}}''',\tau''')\,\rangle_2 \,.
\end{eqnarray}
The trace $\langle \cdots \rangle_2$ is taken over the second order cumulant expansion of the total Hamiltonian.

We specialize the equation (\ref{chicum}) to the case  ${\mathbf{P}}={\mathbf{P}}'$ and $\pmb{\rho}=\pmb{\rho}'$, and apply  Wick's theorem to the trace over $H_0$. Recalling the definition of the dynamic correlation function $\chi$ (Eq. \ref{chifunc}) and of the particle-particle bubble $\mathcal{G}$ (Eq. \ref{gstorta}), we can make the following identification:
\begin{eqnarray}
\langle T\left[ Q^{\dagger}({\mathbf{P}},\pmb{\rho},\tau)Q({\mathbf{P}},\pmb{\rho},\tau')\right]\rangle_2 &=&i\,\chi({\mathbf{P}},\pmb{\rho},\pmb{\rho},\tau-\tau')\\
\langle  T\left[ Q^{\dagger}({\mathbf{P}},\pmb{\rho},\tau)Q({\mathbf{P}},\pmb{\rho},\tau')\right] \rangle_0 &=&-\,{\mathcal{G}}({\mathbf{P}},\mathbf{0},\tau-\tau')\\
 \langle  T\left[ Q^{\dagger}({\mathbf{P}},0,\tau)Q({\mathbf{P}},\pmb{\rho},\tau') \right]\rangle_0\ &=&-\,{\mathcal{G}}({\mathbf{P}},\pmb{\rho},\tau-\tau')\,.
\end{eqnarray}
Performing a Fourier transformation with respect to time,  we finally  recover from equation (\ref{chicum}) the result obtained with a Bethe-Salpeter approach:
\begin{eqnarray}\label{equivalence}
&&\chi({\mathbf{P}},\pmb{\rho},\pmb{\rho},z_{\nu})\,= \nonumber \\
&&=\,-i\int d\tau e^{i\,z_{\nu}\tau}\langle T\left[ Q^{\dagger}({\mathbf{P}},\pmb{\rho},\tau)Q({\mathbf{P}},\pmb{\rho},0)\right]\rangle_2 \nonumber \\
&&= \,-i {\mathcal{G}}({\mathbf{P}},{\mathbf{0}},z_{\nu})\,+\,{\mathcal{G}}({\mathbf{P}},\pmb{\rho},z_{\nu}){\mathcal{G}}({\mathbf{P}},-\pmb{\rho},z_{\nu}) \notag \\
&& \times  \frac{U}{1\,-\,i\,U\,\sum_{{\mathbf{k}}}\,\tilde{{\mathcal{G}}}({\mathbf{P}},{\mathbf{k}};z_{\nu})}\,,
\end{eqnarray}
where the Fourier transform of the trace of the classical fields calculated  with respect to the second order cumulant expansion distribution has been  identified with the $T-$matrix:
\begin{eqnarray}
\int d\tau e^{i\,z_{\nu}\tau}\langle \Delta({\mathbf{P}},\tau) \Delta^*({\mathbf{P}},0)\rangle_2 &= &i\,T({\mathbf{P}},z_{\nu})\nonumber \\
&=&\,\frac{i\,U}{1\,-\,i\,U\,\sum_{{\mathbf{k}}}\,\tilde{{\mathcal{G}}}({\mathbf{P}},{\mathbf{k}};z_{\nu})}\,.\nonumber \\
&&
\end{eqnarray}
Equation (\ref{equivalence}) is equivalent to (\ref{chiTm}) for the case of a local potential (i.e. $\phi_{\lambda}({\mathbf{k}})=1$).

Expression (\ref{equivalence}) calculated for $\pmb{\rho}=\mathbf{0}$ is equivalent to the r.h.s. of equation (\ref{relchidel}). This shows again that the number of Cooper pairs calculated via eq. (\ref{NCP}) gives the number of Cooper pairs with \emph{fixed} internal distance, i.e. zero internal distance for an on-site potential.


\newpage

\end{document}